# Trade-offs in the Design of Multimodal Interaction for Older Adults

Gianluca Schiavo, Ornella Mich, Michela Ferron and Nadia Mana

*FBK - Fondazione Bruno Kessler, Trento, Italy*

**Abstract**

This paper presents key aspects and trade-offs that designers and Human-Computer Interaction practitioners might encounter when designing multimodal interaction for older adults. The paper gathers literature on multimodal interaction and assistive technology, and describes a set of design challenges specific for older users. Building on these main design challenges, four trade-offs in the design of multimodal technology for this target group are presented and discussed. To highlight the relevance of the trade-offs in the design process of multimodal technology for older adults, two of the four reported trade-offs are illustrated with two user studies that explored mid-air and speech-based interaction with a tablet device. The first study investigates the design trade-offs related to redundant multimodal commands in older, middle-aged and younger adults, whereas the second one investigates the design choices related to the definition of a set of mid-air one-hand gestures and voice input commands. Further reflections highlight the design trade-offs that such considerations bring in the process, presenting an overview of the design choices involved and of their potential consequences.

**Introduction**

Multimodal interfaces, meant as "interfaces able to process two or more combined user input modes, such as speech, pen, touch, manual gestures, and gaze, in a coordinated manner with multimedia system output" (Oviatt, 2003, p. 414), seek to combine multiple sensory input and output channels in similar ways as in natural interaction. This similarity has led to the expectation that multimodality in Human-Computer Interaction (HCI) can provide a more natural, robust and flexible form of interaction with respect to more traditional input modalities such as mouse and keyboard (Turk 2014). In this respect, multimodal human–computer interaction has sought to provide not only more powerful and compelling interactive experiences, but also more accessible interfaces to technological devices. Moreover, following the principle of *design for all* and *inclusive design*, multimodal technology has been proposed as a possible solution that allows users to use the interaction modality that best suits their preferences and/or needs, thus making the interaction more flexible. However, despite these potential advantages of multimodal interfaces, the literature reports significant disadvantages as well. For example, different modalities may interfere with each other and a synchronization problem might arise. Additionally, combining and coordinating more than one modality might also require more effort from the users (Naumann, Wechsung, and Hurtienne 2010; Wechsung and Naumann 2008) and a higher cognitive load (Naumann, Wechsung, and Hurtienne 2010). Current research provides findings supporting both assumptions by reporting advantages, as well as disadvantages.

This paper aims to further advance the discussion on this topic by presenting design trade-offs in multimodal technology when designing technology for older adults. In literature, only few works have investigated the design process and design choices for multimodal technology. Here the discussion is based on the analysis of existing guidelines from multimodal interaction and older-adults HCI and User-Centred Design (UCD) literature in order to investigate the effect of different design challenges in multimodal interaction according to four different design trade-offs. In two studies we investigated the multimodal interaction based on the combination of mid-air gestures and vocal commands in order to illustrate two design challenges (*semantic organization* and *interaction saliency*) and two trade-offs related to the design choices (*balancing complexity* and *balancing automation*) that are particularly relevant for mid-air gestural and vocal interaction. These two examples illustrate how design choices should be weighted and how assessing advantages, and corresponding disadvantages, in terms of design trade-offs might support the design process.

This paper is organized as follows: after introducing multimodal interaction and specificities of older users, the design process of multimodal technology for older adults is discussed, with particular attention to the main design challenges and guidelines reported in the literature. Starting from the recommendations identified through the study of the literature, the effect of different design challenges in multimodal interaction in the light of four different design trade-offs are presented. To illustrate the relevance of these trade-offs in the design process, two user studies investigating the design choices related to mid-air gestural and vocal interaction for older adults are presented. In the two user studies two design trade-offs, namely balancing complexity and balancing automation, are explored. Finally, the results of these investigations are discussed and the conclusions drawn.

**Multimodal Interaction for Older Users**

Multimodal interfaces have been considered to improve accessibility for a number of users and usage contexts (Obrenovic, Abascal, and Starcevic 2007), including the diverse needs of older users (Himmelsbach et al. 2015; Munteanu and Salah 2017). Multimodal systems can integrate a wider range of modalities (such as speech, writing, gaze, touch or mid-air gestures) and potentially better accommodate users' preferences with respect to unimodal interfaces. Furthermore, people who have little or no experience with common computer devices can find multimodal interfaces more user-friendly since they offer the possibility to use multiple interaction channels instead of relying on a single source of input (Himmelsbach et al. 2015). However, other studies point out that multimodality must be carefully introduced since it might require more cognitive effort to coordinate different input modalities (especially when more than two modalities are involved) and additional physical demand (Naumann, Wechsung, and Hurtienne 2010). This may become particularly relevant when considering the cognitive and physical characteristics of older users (Fisk 2009).

Numerous examples of multimodal technology for older adults can be found in research and on the market. For instance, social robots or telepresence technology are two representative examples of multimodal systems believed to assist and support older users (Munteanu and Salah 2017). Mobile technology is another field in which multimodal interaction is experimented, given the opportunity that mobile context offers (Lemmelä et al., 2008). Before listing the design trade-offs that multimodal interaction might bring to the technology, we summarize a list of considerations to be taken into account when designing technology for older adults.

**Designing multimodal technology for older adults**

In the present work, we consider older adults those people who are 65 or older (Farage et al. 2012), even if we share the opinion of several authors in believing that grouping older people exclusively by their chronological age is restrictive, since chronologically older adults do not constitute a homogeneous group (Vines et al. 2015). Indeed, they can be very diverse if we take into account their life style and circumstances such as physical condition, cognitive ability, health, income and living arrangements (Lindsay et al. 2012). Having said that, it cannot be denied that ageing brings about several changes covering different aspects of life, such as changes in perception, cognition, movement control, psychological and social well-being, as well as shifts in the social environment and a higher incidence of age-related health problems (Fisk et al. 2009; Farage et al. 2012; Seeman et al. 2001; Hawthorn 2000).

In HCI research studies, and particularly in the design of multimodal technology for older adults, three main aspects should be considered:

**(1) The influence of cognitive factors.** Because of the age-related changes, older adults can be considered a specific user group with respect to younger adults, as they (a) might need more time to learn how to use digital tools, (b) might be more error-prone, and (c) might require a specific kind of support and interface design. Due to short-term memory impairment and lower fluid intelligence, any new system is harder to learn for older people. Several studies (e.g., Venkatesh et al. 2003; Barnard et al. 2013) affirm that a series of high-quality, short, and repetitive training sessions should be provided in order to reinforce the learning of basic commands to operate a new system.

**(2) Physical performance and fatigue.** Older participants can feel fatigued more easily than younger ones, especially when using gestural interaction or moving their upper limbs. Attention should be paid to avoid as much as possible additional risks from injury, pain or fatigue (Gerling, Klauser, and Niesenhaus 2011; Lepicard and Vigouroux 2012).

**(3) Acceptability and long-term use of technology.** Research on technology acceptance has shown that older adults, compared to younger users, decide to adopt new technologies differently. Multiple factors, such as computer-related knowledge, technical self-confidence, previous computer experience, user's performance, the presence of efficient technical support, fear of failure, effective user interaction and usability, concur to form such decision (Wilkowska and Ziefle 2009). These factors were found to be mainly related to ease of use, one of the components of technology acceptance (Davis 1989). In addition to this, new technologies need to satisfy also the second component of technology acceptance, usefulness. To this regard, the lack of perceived advantages may explain the reluctance of many older adults to use novel digital technologies (Melenhorst, Rogers, and Caylor 2001). Indeed, perceived benefits play a significant role in fostering the motivation that leads to the adoption of novel technologies in the long run.

*Guidelines for designing multimodal interaction for older adults*

In this section, we review the literature on multimodal interaction and HCI to specifically investigate trade-offs and frictions in the design of multimodal systems for older adults. This investigation is based on existing research and studies in the field of HCI, UCD and Inclusive Design. Papers were retrieved from ACM digital library, and only articles specifically presenting a summary of guidelines and recommendations for multimodal interaction and for the design of technology for older people were included.

Regarding multimodal interaction, several guidelines for the design of multimodal interfaces for older adults have been discussed in the HCI literature (Oviatt and Cohen 2015; Reeves et al. 2004; McGee-Lennon, Wolters, and Brewster 2011), also considering use cases with older users (Munteanu and Salah 2017; Naumann, Wechsung, and Hurtienne 2010; Xiao et al. 2003).

These studies defined a set of guidelines for multimodal user interface design that are summarized below and reported in Table 1. The rationale behind the guidelines is that each one provides indications or good practices related to a meaningful design problem or challenge that might arise in the process of creating multimodal interaction for older adults.

Table 1. Guidelines for addressing design challenges in designing multimodal interaction for older adults.

| Multimodal design guideline | Design challenge | Interaction context | Related literature |
| --- | --- | --- | --- |
| Give the user or caregiver the choice to select the interaction modality or combination of modalities | **Diverse abilities** | Need to use the most suitable modality | o Munteanu and Salah 2017<br>o Naumann, Wechsung, and Hurtienne 2010<br>o Reeves et al. 2004<br>o McGee-Lennon, Wolters, and Brewster 2011 |
| Consider individual differences in multimodal integration patterns | **Integration patterns** | Need to support user's integration pattern | o Oviatt and Cohen 2015<br>o Xiao et al 2003 |
| Consider the advantages of semantic complementarity or redundancy in the design of multimodal commands | **Semantic organization** | Multiple interaction channels might complement or repeat semantic information | o Oviatt and Cohen 2015<br>o Mills and Alty 1997 |

**Published article:** https://doi.org/10.1080/0144929X.2020.1851768

| Employ well-developed components and rely on complementary modalities to reduce error rates and increase usability | **Technology reliability** | Users need to be able to rely on the technologies | ○ Munteanu and Salah 2017<br>○ Naumann, Wechsung, and Hurtienne 2010<br>○ Reeves et al. 2004 |
|---|---|---|---|
| Combine active and passive triggers. Maintain transparency on how the system works and on how to interact with it. | **Interaction salience** | Need to support transparent, seamless interaction while making the user aware of the data being recorded. | ○ Oviatt and Cohen 2015<br>○ Reeves et al. 2004 |
| The system should dynamically adapt multimodal interfaces to user's preferred or stronger modality | **Adaptation and personalization** | Need to leverage user's strongest or preferred modality | ○ Oviatt and Cohen 2015<br>○ Reeves et al. 2004<br>○ McGee-Lennon, Wolters, and Brewster 2011 |
| The system should support user-initiated interaction, supporting the user to independently interact with the technology. | **Independence** | Support user's need for self-reliance and independence | ○ Munteanu and Salah 2017<br>○ McGee-Lennon, Wolters, and Brewster 2011 |
| Output modalities should respect users' privacy and suit the specific context of use. | **Privacy and context of use** | Multimodality requires specific privacy and contextual requirements | ○ Munteanu and Salah 2017<br>○ Reeves et al. 2004<br>○ McGee-Lennon, Wolters, and Brewster 2011 |

**Diverse abilities.** Multimodal systems should provide users with the choice of the most efficient interaction modality among those proposed by the system. Moreover, users should be able to switch to another interaction modality, for example after a recognition error has occurred in the previous one (Turk 2014). However, this requires that the user knows which is the best modality for her/him, or at least "intuitively" uses the best set of multimodal inputs.

**Integration patterns.** Research on multimodal interaction has shown significant individual differences among how users combine multiple modalities (referred also as multimodal integration patterns, Oviatt, DeAngeli, and Kuhn 1997). There are large individual differences in users' multimodal interaction patterns (Oviatt, DeAngeli, and Kuhn 1997; Xiao et al. 2003; Oviatt, Lunsford & Coulston, 2005): some individuals tend to integrate different modalities simultaneously and overlap them temporally (simultaneous integrators), whereas others tend to completing one mode before starting the next one (sequential integrators). Studies have shown that also older adults demonstrate either a predominantly simultaneous or sequential dominant integration during production of speech and pen multimodal commands (Xiao et al. 2003). Designers should be aware of individual differences in multimodal integration patterns (Oviatt et al., 2005) and multimodal interfaces should be created to accommodate individual interaction patterns.

**Semantic organization.** Complementarity and redundancy are two crucial aspects that should be considered in the design of multimodal interfaces for older adults. Studies have shown the importance of complementarity as an organizational theme in multi-modal interaction (Oviatt and Cohen 2000), while others have highlighted the benefits of redundancy (Mills and Alty 1997), especially when an interaction channel becomes indistinct or noisy. Indeed, a multimodal system can receive redundant information from more than one modality, for instance when a command is given by moving a hand from right to left plus saying "go ahead" in order to select the next item in a horizontal list. This redundancy can support the successful interpretation of the input message by the application, since one stream of information can be used to compensate for the other one during times of distortion or of poor quality. The design trade-offs related to this type of interaction have been further elaborated in the later sections of this article.

**Technology reliability.** Users should be able to rely on multimodal technology, especially in the case of assistive technology. For this reason, multimodal processing should be accurate and robust. However, the fact that recognition algorithms are mainly trained on data from non-older population might pose limitations on the performance of recognition systems due to specific characteristics of older users (e.g., age-related changes on vocal quality that might impact the performance of speech recognition systems (Vacher et al. 2012), or slower gesture speed that might degrade gesture recognition).

**Interaction salience.** Multimodal interaction implies that the user can fluidly switch between the supported input modalities at any time. This implies that the system must be able to seamlessly adapt to the user's interaction, supporting a seamless interaction and limiting the requirement to learn specific commands for interacting with the system. Actually, most multimodal interfaces incorporate trigger mechanisms that activate the interaction with the system when a particular event is detected (e.g., starting speech recognition after the user says "Ok, Google" or "Hey Siri" or when the system senses that the users' lips are moving). Triggers might be active, when they require a direct action from the user (e.g., a trigger word or phrase), or passive, when they are inferred by the system (e.g., lips movement). According to the related literature, the development of multimodal interfaces that rely too heavily on passive triggers, without adequate human control via active input modes, is potentially hazardous and might hinder user experience (Oviatt and Cohen 2015). Among the main issues there are limited system transparency and unintended system consequences due to sensor false activation. On the other hand, active triggers require the user to remember a set of additional commands (and the correct timing for using them) and might hinder the interaction flow. A general recommendation is to maintain transparency on how to interact with the system (Oviatt and Cohen 2000), while at the same time to make both the system's underlying operation and what data is being collected, accessible without being too complex.

**Adaptation and customization.** One-solution-fits-all models are inadequate as they do not consider the characteristics of the individuals. Interaction and interface should be made adaptable and personalized based on user preferences and device characteristics. Many guidelines recommend that users should be able to customize the multimodal channels they would rather use for a given task in an application (Reeves et al. 2004; McGee-Lennon, Wolters, and Brewster 2011). Moreover, studies have found that older adults weigh the trade-offs between modalities differently, and therefore, they should be able to choose from a range of options (McGee-Lennon, Wolters, and Brewster 2011).

**Independence.** Multimodal interfaces should enable older users to independently interact with the technology, even when there is a specific impairment (for example hearing loss or reduced sight). Multimodal interfaces can also contribute to seniors' perceived independence (Munteanu and Salah 2017; McGee-Lennon, Wolters, and Brewster 2011), if they can enable the user to function independently.

**Privacy and context of use.** The context of use should be carefully considered when designing multimodal technology (Neves et al. 2015; Rico and Brewster 2010a, 2010b):

older people have privacy and social acceptability concerns about using some modalities in public spaces (as in the case of voice commands or mid-air gestures (Rico and Brewster 2010a). However, one of the advantages of multimodal interaction is the possibility of using one modality rather than the other according to the specific context (e.g., gestures instead of voice commands in noisy environments).

**Design Trade-Offs in Multimodal Interaction Design for Older Adults**

When considering the guidelines summarized above (Table 1), practitioners and designers might expect to handle a number of design trade-offs (Figure 1), which are situations that involve losing one quality or aspect of the design in return for gaining another quality or aspect (Fischer 2017). This might be particularly relevant when designing technology for a wide range of users, such as people with varying abilities like older adults, or when designing multimodal technology that is based on a wide range of component technologies. In these cases, designers might find themselves taking decisions considering how to apply a specific guideline in the design of multimodal interaction, weighing different options.

In the following, starting from the design challenges identified in the previous section, we discuss the relationship of the different design challenges in multimodal interaction in the light of four different design trade-offs (Table 2).

| Design challenges | Design trade-offs | | | |
|---|---|---|---|---|
| | Balancing complexity | Balancing automation | Balancing personalization | Balancing independence |
| Diverse abilities | ■ | | ■ | ■ |
| Integration patterns | ■ | ■ | ■ | |
| Semantic organization[1] | ■ | ■ | | |
| Technology reliability | ■ | ■ | | ■ |
| Interaction salience[2] | ■ | ■ | | |
| Adaptation and personalization | | ■ | ■ | ■ |
| Independence | ■ | | | ■ |
| Privacy and context of use | | ■ | | ■ |

[1] Design challenge investigated in the first study
[2] Design challenge investigated in the second study

*Figure 1. Design challenges and trade-offs in the design of multimodal interaction for older adults. The user studies here discussed investigate the trade-offs involved in the design challenges of semantic organization (study 1) and interaction salience (study 2).*

**(1) Balancing complexity (Trade-off between complexity and simplicity).** Providing users with the possibility to interact with more than one modality might increase the interaction complexity. This might be true for some combinations of multimodal channels. For example, it has been shown that older adults find some modalities or combination of modalities too complex to use when using multimodal applications (Neves et al. 2015; Naumann, Wechsung, and Hurtienne 2010; Lepicard and Vigouroux 2012). On the other hand, designing interaction with simplicity in mind might force a compromise on functionality (Norman 2010), and this holds true also for multimodal interaction. This trade-off might also affect system usability: a technology that supports many different modalities might increase in complexity and thus be less usable. However, complexity and simplicity are two categories highly investigated in HCI and design research. As others have highlighted (Joshi 2015; Eytam, Tractinsky, and Lowengart 2017), it is important to consider simplicity (and consequently complexity) not as an objective quality, but rather as a quality to be understood through how users perceive simplicity and complexity, thus related to the context of use and level of mastery.

**(2) Balancing automation (Trade-off between automation and control).** Multimodal technology relies largely on the application of advanced sensors and algorithms. In the design of multimodal interaction, attention should be devoted to how to communicate the behaviour of such complex systems to the user, especially when considering older adults (Wu and Munteanu 2018; Broady, Chan, and Caputi 2010). On the one hand, multimodal sensors can be used as background controls, to which the interface automatically adapts without any intentional and direct engagement on the part of the user (Dumas et al., 2013). In this sense, a proactive system might come forward with suggestions, or automatic responses, based on the sensed context and without engaging the user (automation). On the other hand, a reactive system requires the user to initiate action (control), which implies direct attention and focus on the activity.

**(3) Balancing personalization (Trade-off between adaptation and customization).** Multimodal interaction can be tailored to the specific preferences or needs of the user. This process might also end up in an over-personalization of the interaction, making it difficult for the user to discover or experiment with alternative interaction modalities. There is indeed a trade-off between adaptation, where the system adapts the interaction to the user, and customization, where the user is in control of the personalization process. The latter allows users to control the interaction, assuming that they know how and what feature to control. The former gives control to the system without requiring an effort from the user, but it heavily relies on system reliability and performance. Another aspect in multimodal interaction related to this trade-off is the design of multimodal commands: commands can be created by designers and communicated to the user (pre-defined) or they can be defined by the users themselves (user-defined). Considering this distinction, studies in HCI have elaborated on the design trade-offs among different types of sets. As discussed by Nacenta and colleagues (2013), user-defined commands can (a) have a positive effect on accessibility (e.g., people with reduced right-hand mobility could create gestures that do not involve that hand), (b) enable adaptation to the individual's needs, and (c) help leverage people's personal background (e.g., culture, personality, and experiences) to provide easier to remember personal associations. In contrast, pre-defined and stock commands might require users to learn them, but they can be better recognized by the system. Moreover, they can be easily communicated and interpreted by collaborators or other people. However, how the consequences of these design choices when older adults are involved still remain an open question.

**(4) Balancing independence (Trade-off between independence and assistance).** The cognitive effort required from older users to personalize system interaction may be avoided by allowing other users to take care of the process. For instance, a multimodal technology could be designed to be personalized by caregivers or therapists. However, delegating actions to them might further increase their workload and could be perceived as an additional demand or burden. This might also decrease older adults independent use of the technology (Munteanu and Salah 2017).

Table 2. The relation between design challenges and trade-offs. For each challenge a list of questions are reported to help practitioners in making tradeoffs visible in the design process.

| Design challenge | Interaction context | Related design trade-offs | Related questions |
|---|---|---|---|
| *Diverse abilities* | Need to use the most suitable modality | ● *Balancing complexity*<br>● *Balancing personalization*<br>● *Balancing independence* | – *Are the users aware of the most suitable modality?*<br>– *How can the system assist users without being too intrusive?* |
| *Integration patterns* | Need to support user's interaction pattern | ● *Balancing complexity*<br>● *Balancing automation*<br>● *Balancing personalization* | – *How to identify the user interaction patterns?*<br>– *How to support changes in the interaction pattern?*<br>– *Can the system proactively identify the most suitable pattern, or is the user (or caregiver) who can customize it?* |
| *Semantic organization* | Multiple interaction channels might complement or repeat semantic information | ● *Balancing complexity*<br>● *Balancing automation* | – *How do older adults respond to complementary or redundant multimodal commands?*<br>– *How might this influence the interaction experience?*<br><br>[This design challenge and related trade-offs are investigated in the first user study] |
| *Technology reliability* | Users need to be able to rely on their assistive technologies for critical support | ● *Balancing complexity*<br>● *Balancing automation*<br>● *Balancing independence* | – *How can the user recover from errors when the system is not reliable?*<br>– *How to provide sufficient training?*<br>– *Could a simple system be more reliable, but lack some advanced functionalities?* |
| *Interaction salience* | Need to support transparent, seamless interaction while making the user aware of the data being recorded. | ● *Balancing complexity*<br>● *Balancing automation* | – *How is interaction perceived by the older adults?*<br>– *How to support control over the interaction without increasing the system complexity?*<br><br>[This design challenge and related trade-offs are investigated in the second user study] |
| *Adaptation and personalization* | Need to leverage user's strongest or preferred modality | ● *Balancing automation*<br>● *Balancing personalization*<br>● *Balancing independence* | – *Should the system automatically adapt to the user characteristics, or is the user in control of the personalization?*<br>– *How do older adults respond to user-generated or pre-defined commands?* |
| *Independence* | Support user's need for self-reliance and independence | ● *Balancing complexity*<br>● *Balancing independence* | – *Who is in charge of the personalization (the user, the system or a third person, e.g. the caregiver)?*<br>– *How to involve caregivers in the process?* |
| *Privacy and context of use* | Multimodality requires specific privacy and contextual requirements | ● *Balancing automation*<br>● *Balancing independence* | – *How adoption of multimodal interaction is influenced by contextual factors?*<br>– *How older adults interpret and perceive privacy when interacting with multimodal technology?*<br>– *How to support privacy control?* |

The aforementioned list of trade-offs is not supposed to be exhaustive, but it is meant to help practitioners in considering additional factors in the design of multimodal technology with the final goal of making better, or at least more informed, design choices. By answering to the question reported in Table 2, practitioners can clarify trade-off issues and determine if a design is overly ambitious in trying to support competing concerns. To highlight the relevance of the trade-offs in the design process, we present two studies that illustrate the design choices involving the design of multimodal technology for older adults. Specifically, the focus of our research are the two design challenges of semantic organization and interaction salience (Figure 1). These two challenges underline the design trade-offs related to balancing complexity and balancing automation and are highly relevant for multimodal interaction design, especially when considering the initial design phases for defining the interaction vocabulary.

## Case Study: Older Adults Interacting with a Tablet Device through Mid-Air Gestures and Voice Commands

The insights presented in the previous sections are instantiated on a case study, meant as an exploratory work, from a research project on multimodal interaction. The project's goal was to develop multimodal interfaces for mobile devices where the interaction is based on a combination of voice commands and mid-air one-hand gestures, specifically addressing the needs of older adults and visually impaired people. During the course of the project, a number of user studies were conducted to explore how older adults use multimodal interaction when introduced to it for the first time, by investigating their preferences and opinions on different interaction modalities (voice, mid-air gestures, etc.). In order to provide additional insights on the challenges of designing multimodal interaction for older adults, in the following we report on two of the studies conducted within the project, where we focused on the design trade-offs associated to the use of redundant multimodal commands (Study 1) and the trade-off analysis in choosing different sets of multimodal commands (Study 2). In particular, this research addressed the design challenges of Semantic Organization and Interaction Saliency and investigated the associated trade-offs: balancing simplicity vs. complexity and adaptation vs. personalization (see Table 2 and Figure 1).

### *Study 1: Design trade-offs of redundant multimodal commands in older, middle-aged and younger adults*

In a first study we wanted to closely investigate the performance of users interacting with a combination of modalities in a multimodal device, namely mid-air gestures and voice commands. In particular, we wanted to explore whether multimodal interaction characterised by the redundant use of gestures and speech commands varied with participants from different age groups: younger (< 30), middle-aged (45-60) and older adults (> 65).

Redundant multimodal interaction is the use of two (or more) input modes for the same action - for example closing an application saying "close" and waving with the hand. When designing multimodal interaction, redundant commands can be chosen to improve application reliability and robustness, and support the successful interpretation of the input message since two different inputs are available. At the same time, redundant interaction might negatively impact the interaction experience since redundant commands might be perceived as unnecessary, if not detrimental, by the user. In this study, we investigated the trade-off in using redundant commands by comparing performance and opinions of younger, middle-aged and older users.

**Published article:** https://doi.org/10.1080/0144929X.2020.1851768

*Design*

The study used a Wizard-of-Oz (WoZ) approach for investigating how participants use redundant multimodal commands, combining mid-air one-hand gestures with speech inputs to interact with a tablet device (Samsung Galaxy Tab S2 8.0-inch). The study was carried out in a between-subject design, involving participants with different ages: young (25-30), middle-aged (45-60) and older (65-75) adults.

*Participants*

Thirty (30) participants took part in the study, 10 for each age group. Ten older participants were recruited among members of a local senior association; their average age was 68.9 years (SD= 3.62). Middle-aged and younger adults were recruited among the administrative personnel of a non-profit organization; their average age was respectively 51.1 (SD= 2.92) and 30.2 (SD= 3.71) years. All groups included 5 female and 5 male participants. All participants had normal or corrected-to-normal vision, and none of them reported impairments in mobility, in handling objects or in hand/wrist flexibility.

*Procedure*

After a short introduction and training, participants were invited to use the redundant multimodal commands to accomplish the task of taking some pictures with a tablet device. This task was chosen since taking photos with a tablet is nowadays popular (Boulanger, Bakhshi, Kaye, & Shamma, 2016), also among seniors. Furthermore, it is a relatively easy task that can be operated via multimodal interaction. Moreover, daily and recreational use of tablet devices, such as photography, has not received much attention (Carreira et al. 2016), especially when considering older users (Ferron et al., 2019).

Participants were explicitly instructed to hold the tablet device with one hand and to perform the multimodal commands using the other hand and their voice. The '*taking a picture*' task was composed of eight sub-tasks: open the camera application (1), shoot a picture (2), zoom in (3) and out (4) the scene, scroll up (5) and down (6) the effect list, scroll the picture gallery to the right (7) and to the left (8).

During the study session, the "wizard" operated the tablet device to carry out the actions corresponding to the interaction performed by the participants. A facilitator supported the participants through the study procedure by guiding them through the sequence of sub-tasks only if needed. The facilitator would not correct any interaction command, nor provide any feedback to the participants about their interaction. Each interaction performed during the task was video recorded for further video analysis.

*Data analysis*

Data analysis included the analysis of quantitative and qualitative information using a mixed-method approach. Quantitative data were extracted from video analysis of the recorded interaction and were analysed to see how users actually performed the multimodal commands. Each command was coded by interaction type as *gesture-only*, *voice-only* or *multimodal* – where both gesture and voice input were performed. Temporal occurrence was also annotated as *in parallel* if the two modalities were performed with less than 2 sec. delay, otherwise it was considered *in sequence*. Time taken to complete the task was recorded as well. Moreover, at the end of the task, qualitative *individual interviews* were conducted to explore user experience and investigate perception on the redundant commands. During the interview, we invited participants to reflect on the different modalities of interaction they had experienced and to offer suggestions or comments. Questions made during the interview included: "What modality did you prefer?", "Did you prefer to use only one of the two modalities (voice or gesture), or the combination of the two?", "How did you find repeating the commands?"

*Results*

In the video analysis, all interactions performed by the users were annotated. Across all participants, the predominant interaction was multimodal commands (Friedman test: $\chi^2$=53.4, *p*<.01; post-hoc with Bonferroni correction: both *p*<.01). No statistically significant differences were observed between groups. Within the multimodal interactions, hand gestures and speech were frequently performed in parallel ($\chi^2$=46, *p*<.01; post-hoc: both *p*<.01). However, older adults showed fewer parallel interactions compared to the other groups (Kruskal-Wallis (K-W) test: H(2)= 11.1, *p*<.01; post-hoc with Dunn-Bonferroni (D-B) comparisons: *p*<.05 older compared to middle-aged, and *p*<.01 older compared to younger adults), and exhibited more gesture-first interactions (K-W: H(2)= 16.4, *p*<.01; D-B: both *p*<.01).

*Execution time.* Average time for completing the whole task differed between groups (Univariate ANOVA: F(2,27)=9.50, *p*<.01). Older and middle-aged adults were slower compared to younger participants (post-hoc comparisons with Bonferroni correction:

*p*<.05 and *p*<.01 respectively, Figure 2).

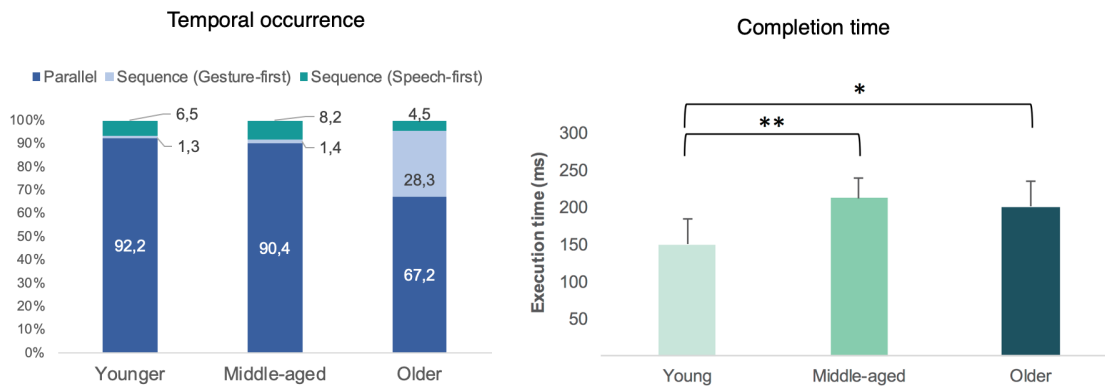

*Figure 2. Temporal occurrence and completion time comparison between younger, middle-aged and older adults (\* p<0.01; \*\*p<0.05).*

*Interviews.* Older adults and the majority of middle-aged adults showed appreciation for multimodal interaction (e.g., Older Participant - OP02: "I liked to take photos in this way", Middle-aged Participant - MP09: "I liked to ask the tablet to shoot a photo for me!"), and preferred voice interaction compared to gesture only (e.g., OP08: "Using the voice is so natural", MP077: "If I have to choose between talking and gesturing, I prefer to use my voice"). Older and middle-aged adults were not concerned about the redundancy of repeating the same command using the gestures and the voice (OP09: "It's natural: when you talk, you use your hands", MP05: "It is hard to make mistakes if I use both my voice and my hands"). The concerns of the older adults about voice interaction focused on the social acceptability of using voice commands in public (e.g., OP01: "I don't want others to hear when I want to take a picture", OP04: "If I have to speak out loud, I will bother the people around me", OP07: "I love to take pictures during my walks. If I am alone I would use the voice, but if I am with my friends, I would prefer to just touch the screen", OP06: "Gesturing to the tablet might be seen as weird if there are other people around"). This aspect is in line with findings on acceptability of multimodal interaction in public places (Rico and Brewster 2010b). Younger adults largely preferred single-modality interaction (especially gestures) or the most common touch interaction (e.g., Younger participant - YP01: "This tablet has been made for being touched"). In particular, they were more negative about speech (e.g., YP09: "I am not used to talking to a tablet", YP06: "It feels awkward to speak out loud to give a command when I can just use my hands") and multimodal interaction (e.g., YP08: "Why should I repeat myself by using both gestures and speech?").

*Discussion*

The findings of this user study point out both similarities and differences between age groups in using redundant multimodal inputs and explore the trade-off of balancing complexity (vs. simplicity) that is related to the use of redundant commands. Overall, multimodal interaction was the predominant type of interaction: all groups, including older participants, were able to easily combine modalities in a redundant way when interacting with the tablet device. We also observed a tendency in older participants to perform mid-air gestures before using the voice commands. When using multimodal commands, older participants used more gesture-first commands compared to the other groups. When older participants were unsure about which word to use, they first performed the gesture and then gave the voice input. This was not the case for middle-aged and younger adults. Furthermore, we also observed that multimodal commands performed by older adults were less synchronous than those performed by the other two groups. In other words, older adults tended to perform more in-sequence commands compared to the other two groups.

In line with previous research on gestural interaction (Stößel and Blessing 2010), older and middle-aged adults were slower than younger participants in performing the commands. In addition, older adults were less concerned about the redundancy of repeating the same command using the gestures and the voice. This might indicate that redundancy might not negatively influence acceptability, at least for older users. However, potential limits related to the social context in which the interaction is performed should be carefully considered in the design of multimodal interfaces, as emerged from the interviews.

All these findings proved useful to inform the design of multimodal commands that were tested in a subsequent user study involving only older participants.

***Study 2: Evaluating mid-air gestures and voice commands sets with older adults***

In a second study, we investigated which of two sets of multimodal commands were preferred by a group of older participants. The two sets differed for the number and type of multimodal commands included (all combining voice commands and mid-air one-hand gestures). The user study tested the two different sets and explored the related interaction experience of a group of twenty older users, exploring the design trade-off associated to the adoption of the two sets.

*Multimodal Command Design*

Two different sets of multimodal commands were compared in the study (Figure 3):

*Complete command set.* The set includes a total of 6 commands, each composed by a mid-air gesture and a vocal command. Four commands were used to navigate the interface (scroll up and down a vertical list, scroll left or right a horizontal list). Two commands were designed for specific interaction not related to spatial navigation, i.e. one for returning to the previous screen ("back") and the other for selecting an element in the screen ("select"). These two commands were designed after an elicitation study conducted with HCI experts and with older adults (Ferron, Mana, and Mich 2019).

*Simple command set.* The simple command set was designed using only 4 commands. Compared to the complete set, the function of the two commands not related to spatial navigation ("back" and "select") were mapped to two navigation commands: returning to the previous screen ("back") was mapped to the "left" gesture, while selecting an item was mapped to the "right" gesture. With this set, items in both vertical and horizontal lists could be scrolled using the up/down gestures.

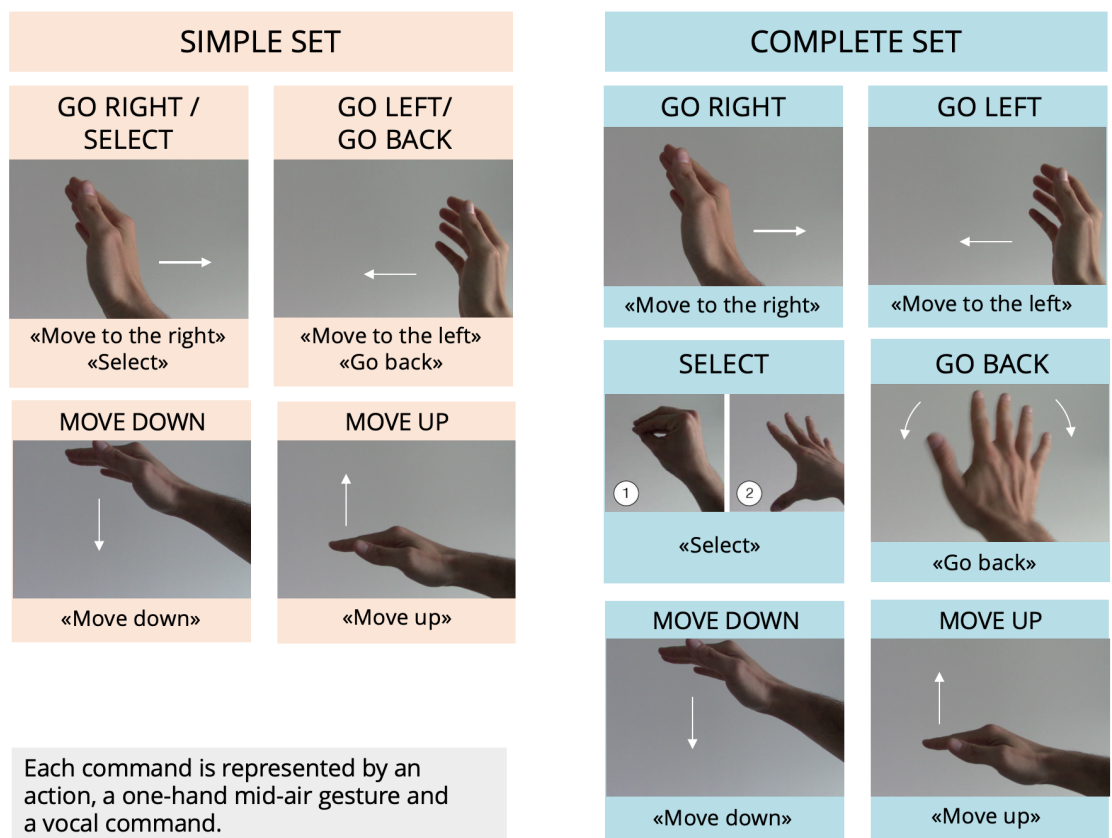

*Figure 3. The two command sets tested in the study.*

The two sets of gestures were analysed considering the trade-off discussed in the previous section, specifically the balance between simplicity and complexity and the trade-off between automation and control (see Table 3).

Table 3. *Analysis of trade-offs between simple and complete command sets.*

|  | *Balancing Complexity* | *Balancing Automation* |
|---|---|---|
| ***Simple set*** | Few commands to remember, but each command has multiple functions. All commands are related to spatial navigation. | The result of each interaction depends on the system state. |
| ***Complete set*** | More commands to remember, each command has one function. Some commands are user-defined. | Having specific commands for each function gives more control over the interaction. |

*Study Design*

The study was carried out in a between-subject design with two groups of 10 participants each (20 in total). Participants were asked to complete two similar tasks (i.e. play a podcast and an audiobook) while interacting with a prototype device that enabled multimodal interaction with the combination of mid-air one-hand gesture and voice commands. Also in this study, the task was chosen to represent daily and recreational use of tablet devices (Carreira et al., 2016).

*Participants*

Twenty (20) older adults (10 Males and 10 Females) took part in the study (Mean age 71 years, SD= 8.1). None of the participants were familiar with the multimodal application and were interacting with it for the first time. Participants were divided into two groups, each group was balanced as for gender composition, age (M= 71 (6.5) and M= 71.1 (9.8), t(18)=0.03, *p*=0.98) and had similar attitudes and habits toward technology (measured through the Attitudes Toward Technologies Questionnaire - ATTQ (Zambianchi and Carelli 2018), M= 3.6 (0.7) and M= 3.2 (1.2), t(18)=0.96, *p*=0.35).

*Material and procedure*

In this study, metrics related to the objective and subjective performance measures of the multimodal system are considered.

Regarding objective performance, recognition accuracy scores were calculated as the percentage of correctly recognised commands over the total number of attempts. This metric assesses the reliability of the system, considered as the number of interaction attempts that were correctly recognised by the multimodal sensor.

Subjective system assessment was measured with a modified version of the SASSI (Hone and Graham 2000) and USE (Lund 2001) questionnaires. SASSI was originally developed to assess users' perception of speech system interfaces and we included four scales in this study (Perceived Accuracy, Cognitive Demand, Annoyance and Speed) that were adapted to cover both speech and mid-air gesture commands. The interaction experience was evaluated using the items from the USE questionnaire, considering four scales related to Usefulness, Ease of Learning, Ease of Use and Satisfaction.

During the study, participants were randomly assigned to one of the two command sets. They first completed a tutorial on the multimodal interaction and then executed the tasks. Participants evaluated the interaction responding to the questionnaire after solving all tasks.

*Results*

On average, recognition accuracy was M= 78% (12) with the simple set and M= 69% (15) with the complete set, and no statistically significant difference was observed (t(18)= 1.6, *p*=0.13). This result indicates that no significant differences exist in the system accuracy with respect to the two sets.

Regarding the questionnaire scores (Figure 4), a statistically significant difference was observed between ratings between the two sets (one-way MANOVA: $F(4,15)=3.04$, $p<0.05$). The complete gesture set scored generally lower with respect to the simple set when considering the user interaction experience (univariate ANOVAs are reported in Table 4). Follow-up analysis revealed that users rated the simple set as easier to use and learn. They were more satisfied with the system and perceived it as more useful compared to the other group. They also perceived the system as more accurate, even though no significant difference between accuracy metrics was observed. No differences were found in the cognitive demand, annoyance and perceived system speed scales.

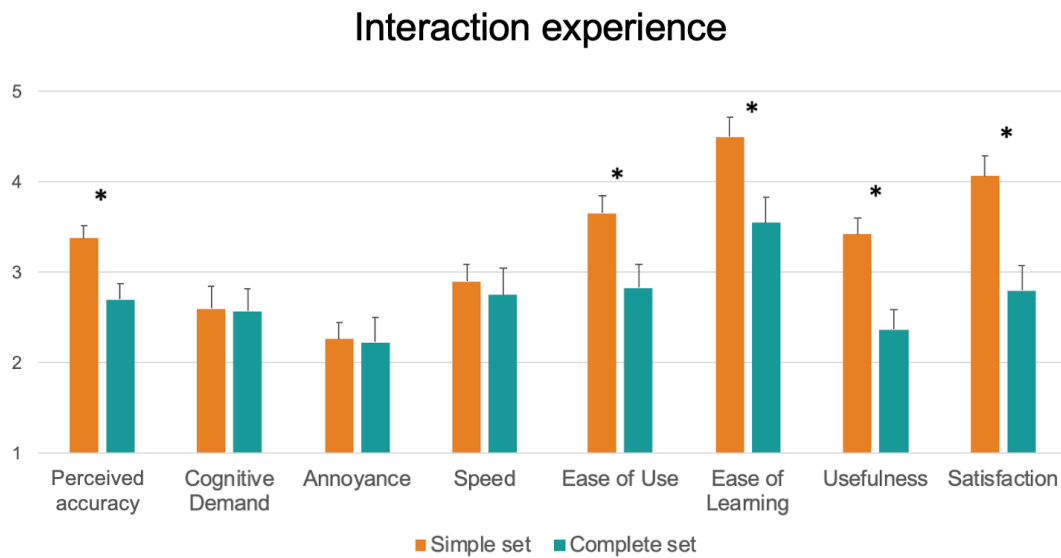

*Figure 4. Mean scores (error bars indicate standard deviations) as reported on the questionnaire scales between the two conditions (simple vs. complete commands set).*

*Table 4. Follow-up univariate ANOVAs on the interaction experience scales.*

| Scales | Simple set | Complete set | | |
| --- | --- | --- | --- | --- |
| | Mean (SD) | Mean (SD) | F | df |
| *Perceived accuracy* | 3.4 (0.4) | 2.7 (0.6) | 8.7* | 18 |
| *Cognitive Demand* | 2.6 (0.8) | 2.6 (0.8) | 0.01 | 18 |
| *Annoyance* | 2.3 (0.6) | 2.2 (0.9) | 0.01 | 18 |
| *Speed* | 2.9 (0.6) | 2.7 (0.9) | 0.2 | 18 |
| *Ease of Use* | 3.7 (0.6) | 2.8 (0.8) | 6.6** | 18 |
| *Ease of Learning* | 4.5 (0.7) | 3.5 (0.9) | 6.9** | 18 |
| *Usefulness* | 3.4 (0.6) | 2.4 (0.7) | 13.9* | 18 |
| *Satisfaction* | 4.1 (0.7) | 2.8 (0.9) | 12.7* | 18 |

\* $p<0.01$; \*\* $p<0.05$

*Discussion*

The results provide evidence of preference toward the use of the simple command set. The trade-offs analysis showed both challenges and opportunities of both commands sets, highlighting the potential consequences related to the use of each set in terms of simplicity and automation. On the one hand, the complete set was perceived as more difficult to learn and use, even though it provided the user with more control over the interaction. On the other hand, the simple set resulted in more positive feedback from the users. Significantly, the system was also perceived as more accurate when using the simple command set, even though no significant differences on recognition accuracy scores were observed between the two conditions.

The results of this second study demonstrate the effect of different design choices and the resulting outcome in terms of trade-offs: the complete set is an option that increased the overall complexity of the system (that was indicated by the lower scores reported by the users), but it might allow more control over the interaction. Conversely, the simple set included a lower number of commands but relied on a higher degree of automatism since the system disambiguated the command function (e.g., "go left" or "back") depending on the interface state (e.g., whether browsing a horizontal list or a vertical menu).

**Conclusion**

This paper has presented some of the challenges and trade-offs that designers, HCI and UCD practitioners might encounter when designing multimodal interfaces. On the one hand, this paper presents a reflection on how to identify the design trade-offs for multimodal interaction for older adults. On the other hand, this work provides practitioners with a framework for analysing and dealing with such trade-offs.

This paper has analysed design trade-offs in the creation of multimodal technology for older adults. Specifically, we identified four main trade-offs related to the balancing of (1) simplicity, (2) automation, (3) personalization and (4) independence. Even though our analysis focuses on older adults as target user group, we believe that most of these trade-offs might also hold for the wider user population.

Discussing the case study of mid-air gesture and vocal interaction, we explored some design trade-offs in the design of such interaction mechanism for older adults in two user studies. The first study investigated the consequences of interacting with multimodal commands by comparing performance and feedback from users of different age groups. We discussed the design trade-off between simplicity and complexity while creating multimodal interaction that requires the use of redundant commands. With this regard, the results from the first study show that older adults, as well as younger participants, could easily combine different interaction modalities in a redundant way, reporting less concern about redundancy with respect to younger users. This suggests that, while redundancy might hinder technology adoption for younger adults, it might not negatively influence acceptability for older adults. Implications for HCI practitioners include the design of temporally adaptive systems, which gradually guides older users through beginners to expert levels. At least initially (and optionally) these systems could comprise redundant multimodal interaction to (a) increase the recognition robustness when users may be more error-prone, (b) compensate for the slower execution of commands by older users, and at the same time (c) foster learning with frequent repetition of gestures and voice commands.

In the second study, we found that a simple set of four multimodal commands received more positive feedback with respect to a larger set of six commands. Even if the complete set allowed more control over the interaction and was richer in terms of semantic interpretation, users preferred the simple set that was also considered easier to learn and to use. This second study showed how design trade-offs might affect the overall user perception about interaction and technology performance. Designers may consider including a simple set of multimodal commands as a default setting, allowing the users the time to get acquainted with the interaction modality and offering the opportunity to optionally choose a more complex set in a later stage. It seems reasonable to expect that once older users familiarize themselves with multimodal interaction, they may opt for a command set that allows more control, rather than for the simple one. However, all the older adults who participated in our studies were novice users with regard to mid-air gestures and voice commands, therefore we were not able to assess the preferences of expert older adults with this respect.

In both user studies our analysis helped to clarify tradeoff issues and identify both acceptable and less acceptable solutions. However, while the framework was successfully used in the two studies, it needs to be further investigated and evaluated.

The user studies covered a particular type of multimodal interaction (the combination of mid-air one-hand gestures and vocal commands) and a specific subset of design trade-offs (mainly the balance between simplicity/complexity and autonomy/control). Future work is needed to deeply explore the influence of certain design choices on different interaction modalities (for example tangible, gestural, and touch-based modalities). Moreover, we believe the list of the trade-offs is not exhaustive and that different aspects can emerge when considering for instance different application domains or different user groups. Furthermore, future work should provide a more in-depth exploration of the design trade-off related to the personalization of a multimodal system and the balance between independence and autonomy in the use of such technology by older adults, also by extending the sample of users in order to overcome this limitation of the first user study.

Lastly, we believe that carefully assessing design ideas that consider such trade-offs might help HCI and UCD researchers in developing multimodal technology that can better accommodate older users' characteristics. In this direction, co-design, end-user involvement and value-centered design approaches (Fischer 2017) can certainly help designers to balance different, and often competing, design choices. In this respect, this work provides a guidance for researchers and practitioners to engage in a structured reflection on the design trade-offs that arise when tackling specific design challenge. As shown in the process reported in the two studies, by identifying the design challenges and exploring the corresponding trade-offs in the design of multimodal interaction, user studies and more direct ways of user involvement can help in assessing advantages, and corresponding disadvantages, that necessarily any design choice would have.

# List of tables

*Table 1. Guidelines for addressing design challenges in designing multimodal interaction for older adults.*

| Multimodal design guideline | Design challenge | Interaction context | Related literature |
|---|---|---|---|
| Give the user or caregiver the choice to select the interaction modality or combination of modalities | **Diverse abilities** | Need to use the most suitable modality | o Munteanu and Salah 2017<br>o Naumann, Wechsung, and Hurtienne 2010<br>o Reeves et al. 2004<br>o McGee-Lennon, Wolters, and Brewster 2011 |
| Consider individual differences in multimodal integration patterns | **Integration patterns** | Need to support user's integration pattern | o Oviatt and Cohen 2015<br>o Xiao et al 2003 |
| Consider the advantages of semantic complementarity or redundancy in the design of multimodal commands | **Semantic organization** | Multiple interaction channels might complement or repeat semantic information | o Oviatt and Cohen 2015<br>o Mills and Alty 1997 |
| Employ well-developed components and rely on complementary modalities to reduce error rates and increase usability | **Technology reliability** | Users need to be able to rely on the technologies | o Munteanu and Salah 2017<br>o Naumann, Wechsung, and Hurtienne 2010<br>o Reeves et al. 2004 |
| Combine active and passive triggers. Maintain transparency on how the system works and on how to interact with it. | **Interaction salience** | Need to support transparent, seamless interaction while making the user aware of the data being recorded. | o Oviatt and Cohen 2015<br>o Reeves et al. 2004 |
| The system should dynamically adapt multimodal interfaces to user's preferred or stronger modality | **Adaptation and personalization** | Need to leverage user's strongest or preferred modality | o Oviatt and Cohen 2015<br>o Reeves et al. 2004 |

| | | | ○ McGee-Lennon, Wolters, and Brewster 2011 |
|---|---|---|---|
| The system should support user-initiated interaction, supporting the user to independently interact with the technology. | **Independence** | Support user's need for self-reliance and independence | ○ Munteanu and Salah 2017<br>○ McGee-Lennon, Wolters, and Brewster 2011 |
| Output modalities should respect users' privacy and suit the specific context of use. | **Privacy and context of use** | Multimodality requires specific privacy and contextual requirements | ○ Munteanu and Salah 2017<br>○ Reeves et al. 2004<br>○ McGee-Lennon, Wolters, and Brewster 2011 |

*Table 2. The relation between design challenges and trade-offs. For each challenge a list of questions are reported to help practitioners in making tradeoffs visible in the design process.*

| Design challenge | Interaction context | Related design trade-offs | Related questions |
|---|---|---|---|
| *Diverse abilities* | Need to use the most suitable modality | ● *Balancing complexity*<br>● *Balancing personalization*<br>● *Balancing independence* | – *Are the users aware of the most suitable modality?*<br>– *How can the system assist users without being too intrusive?* |
| *Integration patterns* | Need to support user's interaction pattern | ● *Balancing complexity*<br>● *Balancing automation*<br>● *Balancing personalization* | – *How to identify the user interaction patterns?*<br>– *How to support changes in the interaction pattern?*<br>– *Can the system proactively identify the most suitable pattern, or is the user (or caregiver) who can customize it?* |
| *Semantic organization* | Multiple interaction channels might complement or repeat semantic information | ● *Balancing complexity*<br>● *Balancing automation* | – *How do older adults respond to complementary or redundant multimodal commands?*<br>– *How might this influence the interaction experience?*<br>[This design challenge and related trade-offs are investigated in the first user study] |

| Technology reliability | Users need to be able to rely on their assistive technologies for critical support | • *Balancing complexity* <br> • *Balancing automation* <br> • *Balancing independence* | – *How can the user recover from errors when the system is not reliable?* <br> – *How to provide sufficient training?* <br> – *Could a simple system be more reliable, but lack some advanced functionalities?* |
|---|---|---|---|
| *Interaction salience* | Need to support transparent, seamless interaction while making the user aware of the data being recorded. | • *Balancing complexity* <br> • *Balancing automation* | – *How is interaction perceived by the older adults?* <br> – *How to support control over the interaction without increasing the system complexity?* <br><br> [This design challenge and related trade-offs are investigated in the second user study] |
| *Adaptation and personalization* | Need to leverage user's strongest or preferred modality | • *Balancing automation* <br> • *Balancing personalization* <br> • *Balancing independence* | – *Should the system automatically adapt to the user characteristics, or is the user in control of the personalization?* <br> – *How do older adults respond to user-generated or pre-defined commands?* |
| *Independence* | Support user's need for self-reliance and independence | • *Balancing complexity* <br> • *Balancing independence* | – *Who is in charge of the personalization (the user, the system or a third person, e.g. the caregiver)?* <br> – *How to involve caregivers in the process?* |
| *Privacy and context of use* | Multimodality requires specific privacy and contextual requirements | • *Balancing automation* <br> • *Balancing independence* | – *How adoption of multimodal interaction is influenced by contextual factors?* <br> – *How older adults interpret and perceive privacy when interacting with multimodal technology?* <br> – *How to support privacy control?* |

*Table 3. Analysis of trade-offs between simple and complete command sets.*

|  | **Balancing Complexity** | **Balancing Automation** |
|---|---|---|
| **Simple set** | Few commands to remember, but each command has multiple functions. All commands are related to spatial navigation. | The result of each interaction depends on the system state. |
| **Complete set** | More commands to remember, each command has one function. Some commands are user-defined. | Having specific commands for each function gives more control over the interaction. |

*Table 4. Follow-up univariate ANOVAs on the interaction experience scales.*

**Published article:** https://doi.org/10.1080/0144929X.2020.1851768

| Scales | Simple set | Complete set | | |
|---|---|---|---|---|
| | Mean (SD) | Mean (SD) | F | df |
| *Perceived accuracy* | 3.4 (0.4) | 2.7 (0.6) | 8.7* | 18 |
| *Cognitive Demand* | 2.6 (0.8) | 2.6 (0.8) | 0.01 | 18 |
| *Annoyance* | 2.3 (0.6) | 2.2 (0.9) | 0.01 | 18 |
| *Speed* | 2.9 (0.6) | 2.7 (0.9) | 0.2 | 18 |
| *Ease of Use* | 3.7 (0.6) | 2.8 (0.8) | 6.6** | 18 |
| *Ease of Learning* | 4.5 (0.7) | 3.5 (0.9) | 6.9** | 18 |
| *Usefulness* | 3.4 (0.6) | 2.4 (0.7) | 13.9* | 18 |
| *Satisfaction* | 4.1 (0.7) | 2.8 (0.9) | 12.7* | 18 |

\* *p<0.01*; \*\**p<0.05*

# List of figures

| Design challenges | Design trade-offs | | | |
| --- | --- | --- | --- | --- |
| | Balancing complexity | Balancing automation | Balancing personalization | Balancing independence |
| Diverse abilities | ■ | | ■ | ■ |
| Integration patterns | ■ | ■ | ■ | |
| Semantic organization[1] | ■ | ■ | | |
| Technology reliability | ■ | ■ | | ■ |
| Interaction salience[2] | ■ | ■ | | |
| Adaptation and personalization | | ■ | ■ | ■ |
| Independence | ■ | | | ■ |
| Privacy and context of use | | ■ | | ■ |

[1] Design challenge investigated in the first study
[2] Design challenge investigated in the second study

*Figure 1. Design challenges and trade-offs in the design of multimodal interaction for older adults. The user studies here discussed investigate the trade-offs involved in the design challenges of semantic organization (study 1) and interaction salience (study 2).*

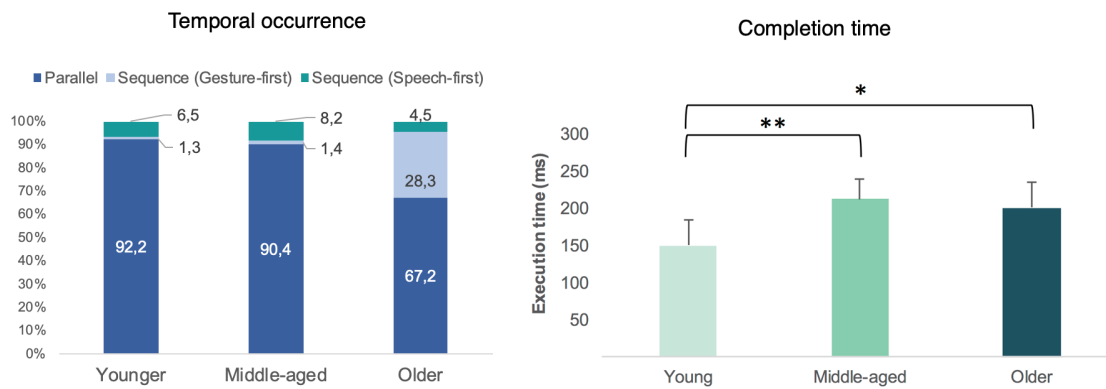

*Figure 2. Temporal occurrence and completion time comparison between younger, middle-aged and older adults (\* p<0.01; \*\*p<0.05).*

Published article: https://doi.org/10.1080/0144929X.2020.1851768

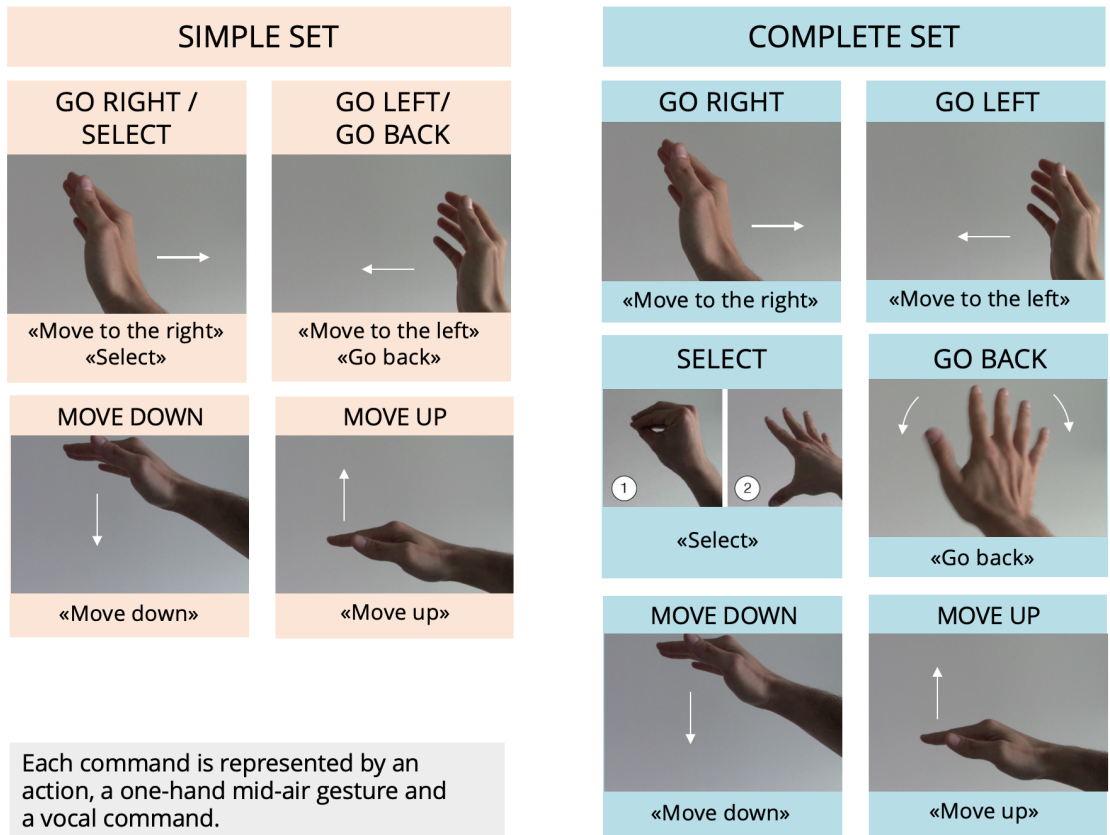

*Figure 3. The two command sets tested in the study.*

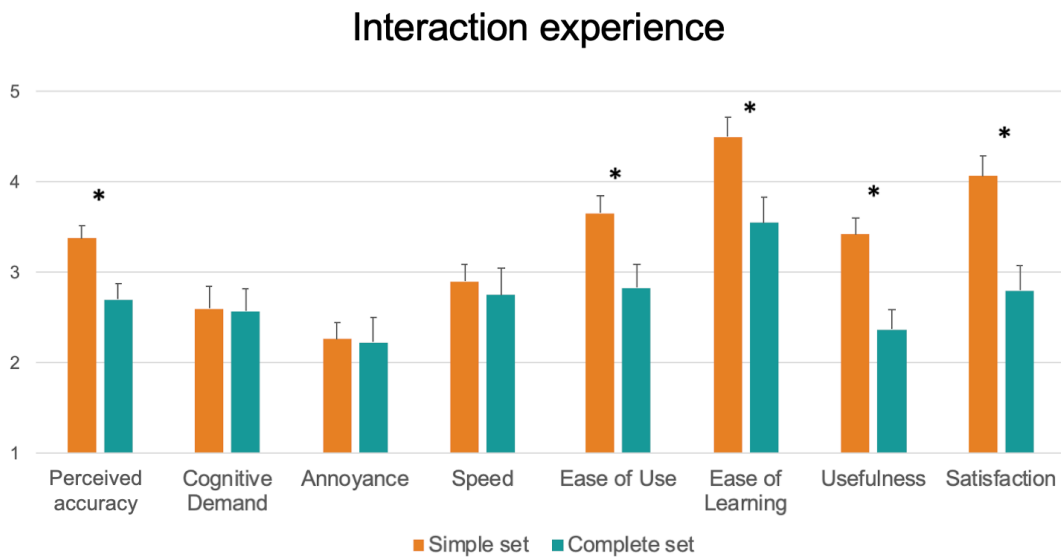

*Figure 4. Mean scores (error bars indicate standard deviations) on the questionnaire scales between the two conditions (simple vs. complete commands set)*

**Published article:** https://doi.org/10.1080/0144929X.2020.1851768